\documentclass[prb,aps,twocolumn,draft,tightenlines,%
showpacs,floatfix,superscriptaddress]{revtex4}
\usepackage{psfig}
\usepackage{amssymb}
\begin{document}
\title{Rashba-effect-induced spin dephasing in $n$-typed InAs quantum wells}
\author{M. Q. Weng}%
\author{M. W. Wu}%
\thanks{Author to whom correspondence should be addressed}%
\email{mwwu@ustc.edu.cn}%
\affiliation{Structure Research Laboratory, University of Science \&%
Technology of China, Academia Sinica,  Hefei, Anhui, 230026, China}
\affiliation{Department of Physics, University of Science \&%
Technology of China, Hefei, Anhui, 230026, China}%
\altaffiliation{Mailing Address.}
\date{\today}
\begin{abstract}
We perform a many-body investigation of the spin dephasing 
in $n$-typed InAs quantum wells 
under moderate magnetic fields in the Voigt configuration 
by constructing and numerically solving the
kinetic Bloch equations. We obtain the spin dephasing time due to the
Rashba effect 
together with the spin conserving scattering such as the electron-phonon, the
electron-nonmagnetic impurity as well as the electron-electron Coulomb
scattering. By varying the initial spin polarization, temperature,
impurity density, applied magnetic field and the interface electric
field, we are able to study the spin dephasing time under various
conditions. For the electron density and quantum well width we study, the many-body
effect dominates the spin dephasing. Moreover, we find an anomalous
resonance peak in the spin dephasing time for high  initial spin
polarization under moderate magnetic fields. 

\end{abstract}

\maketitle

\section {Introduction}

Almost all of the current semiconductor devices are based on
manipulating electronic charges. The arising field of spintronics
proposes to use the spin degree of freedom of electrons in place of/in
addition to the charge degree of freedom for device applications in
order to add new features and functionalities to semiconductors
devices.\cite{spintronics,wolf_sci_2001,spintronics_awsch} 
The hope of the realization of the proposed spintronic devices is supported by
the resent development of ultrafast nonlinear optical
experiments\cite{damen,wagner,baumberg_1994_prl,baumberg_1994_prb,%
heberle,buss1,crooker_1996,crooker_1997,buss2,kikkawa1,kikkawa2,%
kikkawa3,ohno1,ohno}
where long spin dephasing time ($>100$\ ns) is reported. 

The functionalities of the semiconductor spintronic devices rely on the
manipulation of the spin coherence. In order to realize
these devices, one needs to thoroughly understand the spin dephasing
mechanisms which tend to destroy the spin coherence.  
Historically, three spin dephasing mechanisms have been proposed 
in semiconductors:\cite{meier,aronov}
the Elliot-Yafet (EY) mechanism,\cite{yafet,elliot} the D'yakonov-Perel'
(DP) mechanism,\cite{dp} and the Bir-Aronov-Pikus (BAP)
mechanism.\cite{bap} 
All of the three mechanisms are either due to the spin flip (SF)
scattering or are treated as effective SF scattering.  
The spin dephasing times of these mechanisms 
for low polarized system are calculated in
the framework of single particle approximation.\cite{meier} 
In additional to these single particle spin dephasing mechanisms,
three years ago Wu 
proposed a many-body spin dephasing mechanism which has long been
overlooked in the literature. This mechanism is caused by
irreversibly disrupting the phases between spin dipoles due to the 
inhomogeneous broadening together with the spin-conserving (SC)
scattering,\cite{wu_pss_2000,wu_ssc_2002,wu_jpsj_2001,%
wu_epjb_2000,wu_js_2001} and is therefore a many-body effect. 
The inhomogeneous broadening 
can be introduced by the energy dependence of
$g$-factor\cite{wu_pss_2000,wu_js_2001,bronold}
and/or the momentum ${\bf k}$-dependence of 
the DP term.\cite{wu_js_2001,wu_jpsj_2001,wu_ssc_2002} 
Our recent works further show that this 
mechanism also plays an important role in the spin dephasing during the
spin transport.\cite{weng_prb_2002,weng_jap_2003}  

Very recently we performed a systematic investigation\cite{c0302330,c0210313}  
of the spin dephasing due to the DP effect in $n$-typed GaAs (100) quantum wells for high temperatures
($\geq 120$~K) under magnetic fields in the
Voigt configuration by constructing and numerically solving the kinetic
Bloch equations.\cite{wu_prb_2000,%
wu_pss_2000,wu_ssc_2002,wu_jpsj_2001,wu_epjb_2000,wu_js_2001} 
In these studies, we include all the SC scattering 
such as the electron-phonon, the
electron-nonmagnetic impurity as well as the electron-electron Coulomb
scattering and investigate the spin dephasing under various conditions. 
The dephasing obtained from our theory contains both the
single-particle dephasing caused by the effective 
SF scattering first proposed by D'yakonov and Perel',\cite{dp} and
the many-body one due to the inhomogeneous broadening
provided by the DP term. We show that for the electron densities we
studied, the spin dephasing rate is dominated by the many-body effect. 
Moreover, as we include the electron-electron
Coulomb scattering, we are able to investigate the
spin dephasing with extra large spin polarization (up to 100~\%) which
has not been discussed both theoretically and experimentally. 
We find that under moderate magnetic fields, the SDT increases
dramatically with the initial spin polarization. For example, the SDT
of a impurity free sample gets an increase of more than one order of magnitude when
the initial spin polarization rises from about 0 to about 100~\% at low
temperature.\cite{c0302330} 
The initial-spin-polarization dependence of the spin
dephasing becomes more interesting when the magnetic field is increased to
a few tens tesla where the SDT 
no longer increases monotonically with the initial spin polarization
but shows an anomalous resonance peak versus the initial spin
polarization.\cite{c0210313} The dramatic increase and the
anomalous resonance of SDT in the high spin polarization region is
found to be due to the first order of the electron-electron interaction, {\em i.e.}, the Hartree-Fock (HF) 
contribution which provides
an effective magnetic field that can reduce the spin
dephasing and result in a fast increase of the SDT. 
Moreover, under right condition, the HF term, the applied magnetic
field as well as the DP term can reach to a resonance, and thus forms
the anomalous peak. Due to the small Land\'e $g$-factor in GaAs, the resonance condition can only
be achieved under very high magnetic fields.

In this paper, we apply the kinetic theory to study the spin dephasing
in the $n$-typed InAs QW for high temperatures where the DP term is the leading dephasing mechanism.
 In QW system, the DP term is composed of the Dresselhaus
term\cite{dress} and the Rashba term.\cite{ras,rashba} The Dresselhaus
term is due to the lack of inversion symmetry in the zinc-blende
crystal Brillouin zone and is sometimes referred to as bulk inversion
asymmetry (BIA) term.  
Whereas the Rashba term appears if the
self-consistent potential within a QW is asymmetric along the growth
direction and is therefore referred to as structure inversion
asymmetry (SIA) contribution. For QW's of  wide band-gap semiconductors
such as GaAs, the Dresselhaus term is the main spin dephasing mechanism. Whereas for QW's composed of
narrow band-gap semiconductors  such as InAs in the present case, the Rashba term is dominant.
As the Rashba term is proportional to
the interface electric field of the QW, therefore, the spin dephasing
in the InAs QW can be manipulated through applying an electric field
perpendicular to the QW. Moreover, as the Land\`e $g$-factor in InAs is
very large ($g=15$ compared to 0.44 of GaAs), one expects to achieve the resonance condition under a
moderate magnetic field. 
We organize the paper as follows: We present our model and the kinetic
equations in Sec.\ II. Then in 
In Sec.\ III(A) we investigate how the SDT changes with the variation
of the initial spin polarization. The temperature dependence of the
SDT under different spin polarization is discussed in detail 
in Sec.\ III(B), where we also highlight
the difference between the present many-body theory and the earlier
simplified theory. In Sec.\ III(C) we show the magnetic field 
dependence of the SDT. Finally we discuss how the interface electric
field affects the SDT. We present the conclusion and summary in Sec.\ IV.

\section {Kinetic Equations}

We start our investigation from an $n$-doped (100) InAs QW with 
well width $a$. The growth direction is assumed to be $z$-axis. 
A moderate magnetic field {\bf B} is applied along the $x$ axis. 
Due to the confinement of the QW, the
momentum states along $z$ axis are quantized.  Therefore the electron
states are characterized by a subband index $n$ and a two dimensional
wave vector ${\bf k}=(k_x, k_y)$ together with a spin index $\sigma$.
In the present paper, we choose the electron density so that  only the lowest subband 
is populated and the transition
to the upper subbands is unimportant. Therefore, one only needs to
consider the lowest subband.  With the DP term (specifically the Rashba term)
included, the Hamiltonian of the electrons in the QW takes the form:
\begin{equation}
  H=\sum_{{\bf k}\sigma\sigma^{\prime}}\biggl\{
\varepsilon_{\bf k}+\bigl[g\mu_B{\bf B}+{\bf h}({\bf k})\bigr]
\cdot{\vec{\bf \sigma}_{\sigma\sigma^{\prime}}\over 2}\biggr\}
c^{\dagger}_{{\bf k}\sigma}c_{{\bf k}\sigma^{\prime}}+H_I.
\label{eq:hamiltonian}
\end{equation}
Here $\varepsilon_{{\bf k}}={\bf k}^2/2m^{\ast}$ is the energy of
electron with wavevector ${\bf k}$ and effective mass $m^{\ast}$.
$\vec{\bf \sigma}$ are the Pauli matrices. 
The Rashba term ${\bf h}({\bf k})$ can be written as
\begin{equation}
  h_x({\bf k})=\alpha k_y,\;
  h_y({\bf k})=-\alpha k_x,\;
  h_z({\bf k})=0.
  \label{eq:rashba}
\end{equation}
In these equations, $\alpha$ is proportional to the interface electric
field $E_z$ along the growth direction:
\begin{equation}
  \label{eq:alpha}
  \alpha=\alpha_0 eE_z,
\end{equation}
with the coefficient $\alpha_0$ being inversely proportional to the
energy gap and the effective mass.\cite{lommer} 
The interaction Hamiltonian $H_I$ is composed of Coulomb interaction
$H_{ee}$, electron-phonon interaction $H_{ph}$, as well as
electron-impurity scattering $H_i$. Their expressions can be found in
textbooks.\cite{haug,mahan} 

We construct the kinetic Bloch equations by the nonequilibrium Green
function method\cite{haug} as follows:
\begin{equation}
  \label{eq:bloch}
  \dot{\rho}_{{\bf k},\sigma\sigma^{\prime}}
  =\dot{\rho}_{{\bf k},\sigma\sigma^{\prime}}|_{\mbox{coh}}
  +\dot{\rho}_{{\bf k},\sigma\sigma^{\prime}}|_{\mbox{scatt}}
\end{equation}
Here $\rho_{{\bf k}}$ represents the single particle density
matrix. The diagonal elements describe the electron distribution
functions $\rho_{{\bf k},\sigma\sigma}=f_{{\bf k}\sigma}$. The
off-diagonal elements $\rho_{{\bf k},{1\over
    2}-{1\over2}}\equiv\rho_{{\bf k}}$ describe 
the inter-spin-band polarizations
(coherence) of the spin coherence.\cite{wu_prb_2000} Note that 
$\rho_{{\bf k},-{1\over 2}{1\over 2}}\equiv \rho^{\ast}_{{\bf k},{1\over
    2}-{1\over 2}}=\rho^{\ast}_{{\bf k}}$. Therefore, $f_{{\bf
    k}\pm{1\over 2}}$ and $\rho_{{\bf k}}$ are the quantities to be
determined from Bloch equations. 

The coherent part of the equation of motion for the electron
distribution function and the spin coherence are given by 
\begin{widetext}
\begin{equation}
  \label{eq:f_coh}
  {\partial f_{{\bf k},\sigma}\over \partial t}|_{\mbox{coh}}=
-2\sigma\bigl\{[g\mu_BB+h_x({\bf k})]\mbox{Im}\rho_{{\bf k}}+h_y({\bf k})
\mbox{Re}\rho_{{\bf k}}\bigr\}
+4\sigma\mbox{Im}\sum_{{\bf q}}V_{{\bf q}}\rho^{\ast}_{{\bf k}+{\bf
    q}} \rho_{{\bf k}},
\end{equation}
\begin{equation}
  \label{eq:rho_coh}
  {\partial \rho_{{\bf k}}\over \partial t}\left |_{\mbox{coh}}\right. =
  {1\over 2}[ig\mu_B B + ih_x({\bf k}) + h_y({\bf k})]
  (f_{{\bf k}{1\over 2}}-f_{{\bf k}-{1\over 2}})
  +i\sum_{{\bf q}}V_{\bf q}\bigl[(f_{{\bf k}+{\bf q}{1\over 2}}
  -f_{{\bf k}+{\bf q}-{1\over 2}})\rho_{{\bf k}}
  -\rho_{{\bf k}+{\bf q}}(f_{{\bf k}{1\over 2}}
  -f_{{\bf k}-{1\over 2}})\bigr],
\end{equation}
\end{widetext}
respectively, 
where $V_{{\bf q}}=4\pi e^2/[\kappa_0(q+q_0)]$ is the 2D Coulomb
matrix element under static screening. 
$q_0= (e^2m^{\ast}/\kappa_0)\sum_{\sigma}f_{{\bf k}=0,\sigma}$  and
$\kappa_0$ is the static dielectric constant. 
The first term
on the right hand side (RHS) of Eq.~(\ref{eq:f_coh}) describes the spin
precession of electrons under the magnetic field ${\bf B}$ as well as
the effective magnetic field ${\bf h}({\bf k})$ due to the Rashba
effect. The scattering terms of the electron distribution function and
the spin coherence are given by Eqs. (\ref{eq:f_scatt}) and
(\ref{eq:rho_scatt}) in the Appendix.

The initial conditions are taken at $t=0$ as: 
\begin{equation}
\rho_{\bf k}|_{\rm t=0} = 0
\label{eq:rho_init}
\end{equation}
\begin{equation}
f_{{\bf k}\sigma}|_{\rm t=0} = 1/\bigl\{\exp[(\varepsilon_{\bf
  k}-\mu_{\sigma})/k_BT]+1\bigr\} 
\label{eq:fk_init}
\end{equation}
where $\mu_\sigma$ is the chemical potential for spin $\sigma$. The condition
$\mu_{\frac{1}{2}}\neq\mu_{-\frac{1}{2}}$ gives rise to the imbalance
of the electron densities of the two spin bands. 

\section {Numerical Results}

The kinetic Bloch equations form a set of nonlinear equations. All the
unknowns to be solved appear in the scattering terms. Specifically,
the electron distribution function is no longer a Fermi distribution
because of the existence of the anisotropic Rashba term ${\bf h}({\bf
  k})$. This term in the coherent part drives the electron
distribution away from an isotropic Fermi distribution. The
scattering term attempts to randomize electrons in ${\bf
  k}$-space. Obviously, both the coherent part and the scattering
terms have to be solved self-consistently to obtain the distribution
function and the spin coherence. 

We numerically solve the kinetic Bloch equations in such a
self-consistent fashion to study the spin precession between the
spin-up and -down bands. We include electron-phonon scattering
and the electron-electron interaction throughout our computation. 
As we concentrate on the relatively high
temperature regime in the present study, for electron-phonon
scattering we only need to include electron-LO phonon
scattering. Electron-impurity scattering is sometimes excluded. 
As discussed in the previous paper,\cite{wu_prb_2000,kuhn}
irreversible 
spin dephasing can be well defined by the slope of the envelope of the
incoherently 
summed spin coherence $\rho(t)=\sum_{{\bf k}}|\rho_{{\bf k}}|$. 
 The material parameters of InAs for our calculation are
tabulated in Table~\ref{table1}.\cite{made}
The method of the numerical calculation
has been laid out in detail in the previous paper on the
DP mechanism in 3D systems.\cite{wu_pss_2000}
The difference is that here we are able to get the results
quantitatively in stead of only qualitatively as in the previous 3D case,
thanks to the smaller dimension in the momentum space.
Our main results are
plotted in Figs.~\ref{fig3} to \ref{fig8}. 
In these calculations the total electron density $N_e$ 
and the applied magnetic field $B$ are chose to be $4\times
10^{10}$~cm$^{-2}$ and $1.5$~T respectively unless otherwise specified. 

\begin{table}[htbp]
  \centering
  \begin{tabular}{lllllll}
\hline\hline
 $\kappa_\infty$ & \mbox{}\hspace{1.25cm} &
 12.25 & \mbox{}\hspace{1.25cm} &
 $\kappa_0$ & \mbox{}
 \hspace{1.25cm} & 15.15\\ 
 $\omega_0$ & & 27~meV & & $m^*$ & &0.00239~$m_0$\\
$a$&& 7.5\ nm&&&
\\ \hline\hline
  \end{tabular}

  \caption{Parameters used in the numerical calculations}
  \label{table1}
\end{table}

\subsection{Spin polarization dependence of the spin dephasing time}

We first study the spin polarization dependence of the SDT.
As our theory is a many-body theory and 
we include all the scattering, especially the
Coulomb scattering, in our calculation, we are able to calculate the 
SDT with large spin polarization. 

\begin{figure}[htb]
  \psfig{figure=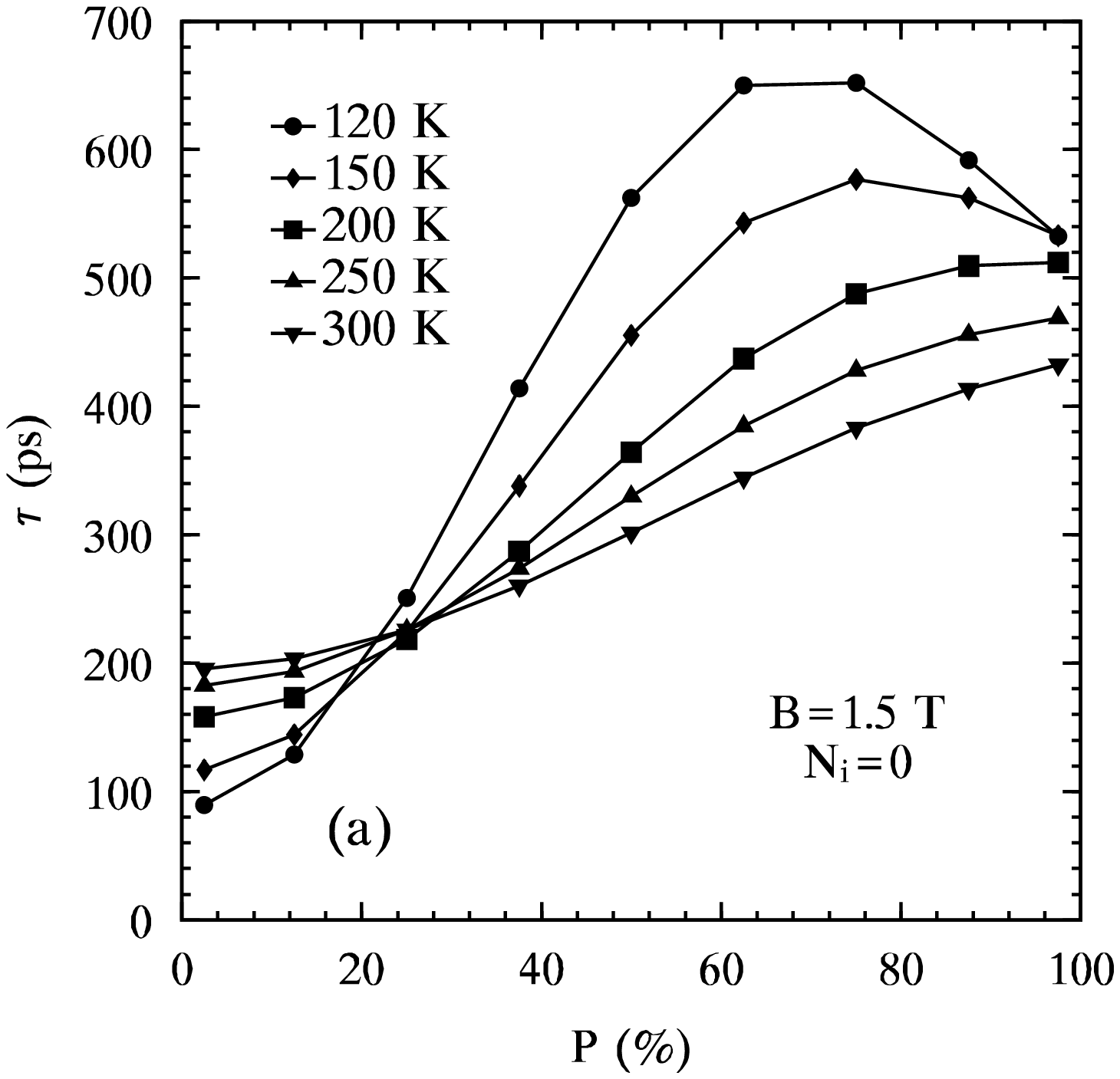,width=9.cm,height=8.5cm,angle=0}
  \psfig{figure=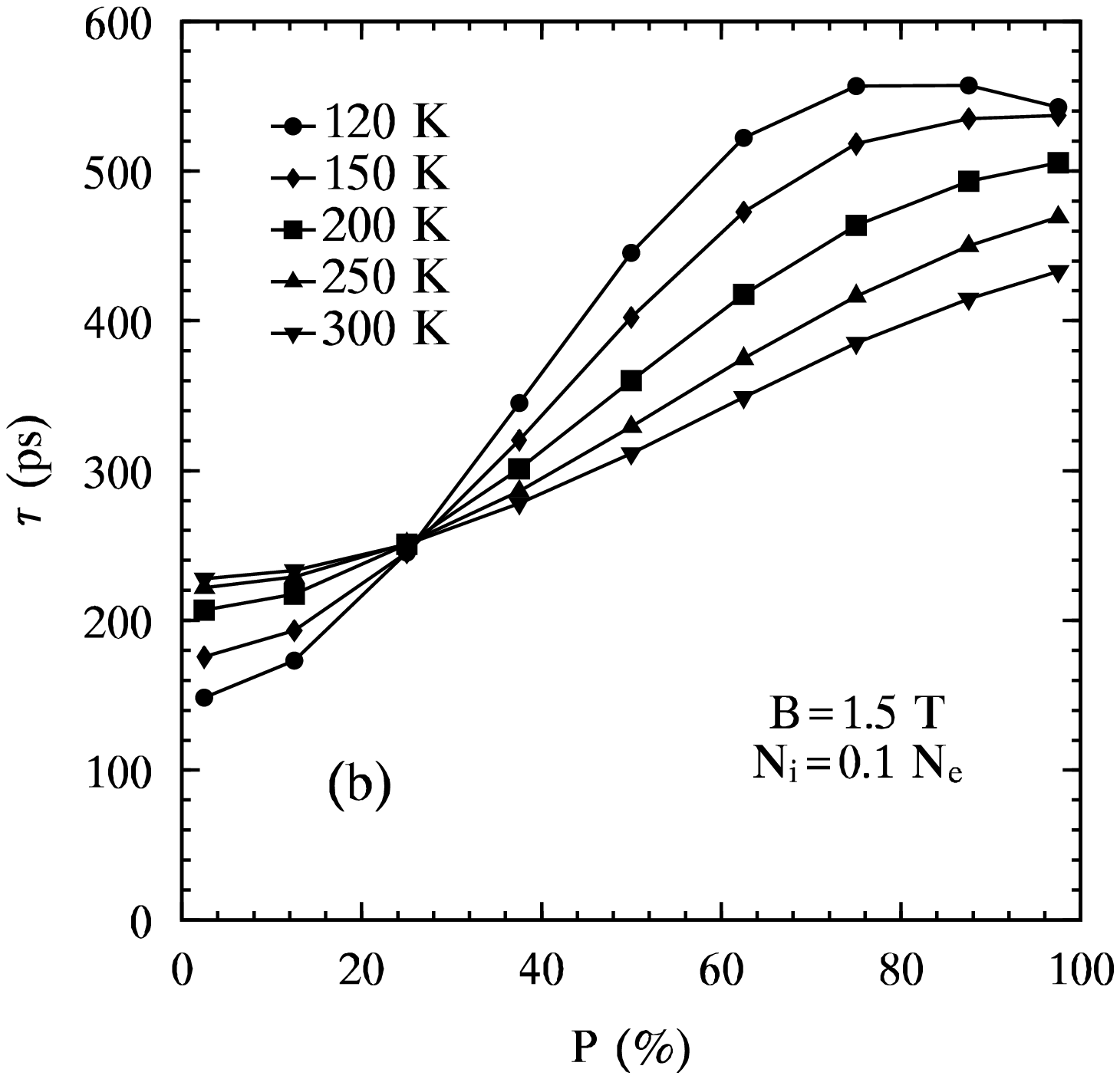,width=9.cm,height=8.5cm,angle=0}
  \caption{Spin dephasing time $\tau$ versus the initial spin
    polarization $P$ 
    with
    different impurity concentration and different temperatures. 
    The impurity densities in (a) and
    (b) are 0 and $0.1 N_e$ respectively. The lines are plotted for
    the aid of eyes. 
  }
  \label{fig3}
\end{figure}

In Fig.~\ref{fig3}, SDT $\tau$ is
plotted against the initial spin polarization $P$ 
with $N_i=0$ [Fig.~\ref{fig3}(a)] and $N_i=0.1 N_e$ [Fig.~\ref{fig3}(b)]
at different temperatures. The most striking feature of the impurity-free
case is the huge anomalous peaks of the SDT in low temperatures.
For  $T=120$\ K, the peak value of the SDT is about 6 times higher
than that of low initial spin polarization. It is also seen from
the figure that the anomalous peak is reduced with the increase of 
temperature and the peak shifts to higher polarization. For $T>200$\ K
there is no anomalous peak.

The anomalous peak in the $\tau$-$P$ curve in low temperature region
originates from the electron-electron interaction, specifically the
Hartree-Fock (HF) self-energy [{\em i.e.}, the last terms in the
Eq. (\ref{eq:f_coh}) and (\ref{eq:rho_coh})].  If one removes the HF
term, the anomalous peak as well as the large increase of SDT
disappears.  It is pointed out in our previous paper that although the
HF term itself does not contribute to the spin dephasing
directly,\cite{wu_epjb_2000,wu_js_2001} it can alert the motion of the
electrons as 
it behaves as an effective magnetic field ${\bf B}^{\mbox{HF}}({\bf
k})$. Therefore, the HF term can affect the spin dephasing by
combining with the DP term.  For small spin polarization as commonly
discussed in the literature, the contribution of the HF term is
marginal. However, when the polarization gets higher, the HF
contribution becomes larger.  Especially the effective magnetic field
formed by the HF term contains a longitudinal component
[$B_z^{\mbox{HF}}({\bf k})$] which can effectively reduce the
``detuning'' of the spin-up and -down electrons, and thus strongly
reduces the spin dephasing, therefore the SDT increases with initial
spin polarization.\cite{c0302330}
Moreover, besides the initial polarization,
$\rho_{\bf k}$ and therefore $B^{\mbox{HF}}({\bf k})$
are also affected by the applied magnetic field.
With higher magnetic
field, both gets larger.  Under the high magnetic field and when the
initial spin polarization reaches to a right value, the effective
magnetic field ${\bf B}^{\mbox{HF}}({\bf k})$ may reach the magnitude
comparable to the contribution from the DP term as well as the applied
magnetic field in the coherent parts of the Bloch equations and
reduces the anisotropic caused by the DP term. Therefore, one gets
much longer SDT.  However, if one further increases the initial
polarization, the HF term exceeds the resonance condition. As the
result, the SDT decreases.  Therefore, one gets the anomalous peak
which is similar to the resonance effect. It is noted at, as both the
DP term and the HF term are $k$-dependent, the resonance is broadened.

For high  temperatures the HF term is smaller. 
In order to reach the resonance, one needs to go to
higher polarization. Therefore, as shown in the figure
the  anomalous  peak shifts to the 
higher polarization. However, when the temperature is high enough,
even largest polarization $P=100$\ \% cannot make the HF term 
to reach the resonance condition. Therefore, the peak disappears.

The $\tau$-$P$ curve is much different when the impurities are
introduced. It is seen from Fig.~\ref{fig3}(b) that, when the
density of impurity is large, say $N_i=0.1 N_e$, the fast rise
in $\tau$-$P$ curve remains.
Nevertheless the increase 
is smaller than the corresponding one when the impurities
are absent. In addition to the reduction of the
rise in $\tau$-$P$ curves, the impurities destroy the anomaly too. 
One can easily see that, with the impurity level $N_i=0.1 N_e$, 
when the temperature is 120~K, the anomalous peak is much flatter than
the impurity free sample, while 
for all other temperatures we study, the SDT
increases uniquely with the polarization. 
\begin{figure}[htb]
  \psfig{figure=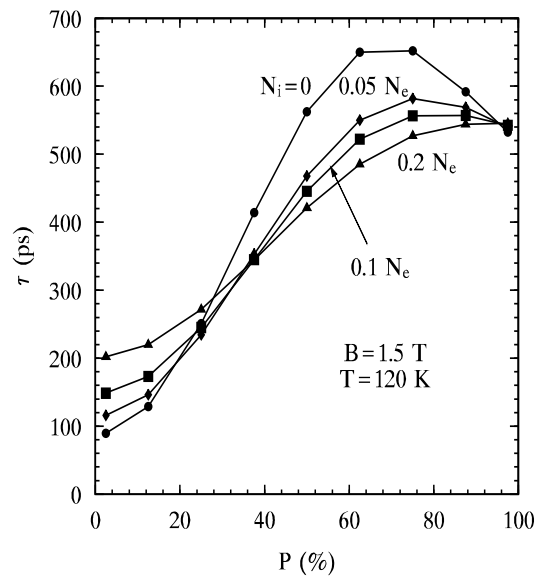,width=9.cm,height=8.5cm,angle=0}
  \caption{Spin dephasing time $\tau$ versus the initial spin
    polarization $P$ for a InAs QW with different impurity levels. 
    Circle ($\bullet$): $N_i=0$;
    Diamond ($\blacklozenge$): $N_i=0.05N_e$;
    Square ($\blacksquare$): $N_i=0.1N_e$;
    Up triangle ($\blacktriangle$): $N_i=0.2N_e$.
    The lines are plotted for
    the aid of eyes. 
  }
  \label{fig4} 
\end{figure}

To further reveal the contribution of the impurity to the dephasing under
different conditions, we plot the  SDT 
as a function of the polarization for different impurity levels at 
$T=120$\ K in Fig.~\ref{fig4}. 
The figure clearly shows that 
the impurity
tends to remove the anomalous peak and to shift the peak to
the larger initial spin polarization. This is because that 
the impurity reduces the HF term and therefore the resonance effect
is also reduced. Hence, in order to reach the maximum resonance, 
one has to increase the initial spin polarization. Consequently, the
peak shifts to larger $P$. 
Whereas when $N_i$ is raised to $0.2 N_e$, 
the HF term is reduced too much to form a peak.

\subsection{The temperature dependence of the spin dephasing time}

Above we discussed the dependence of spin dephasing on
initial spin polarization for different temperatures. 
Now we turn to the temperature dependence of
the SDT under different initial spin polarizations. 
>From Fig.~\ref{fig3}(a) and (b) in Sec.\ III(B),
one can see that for small polarization, 
the SDT increases with the temperature. Whereas in high
polarized region, the SDT decreases with the temperature. For moderate
polarization, the temperature dependence  is too complicated 
to be described by a monotonic function of temperature. 

\begin{figure}[htb]
  \psfig{figure=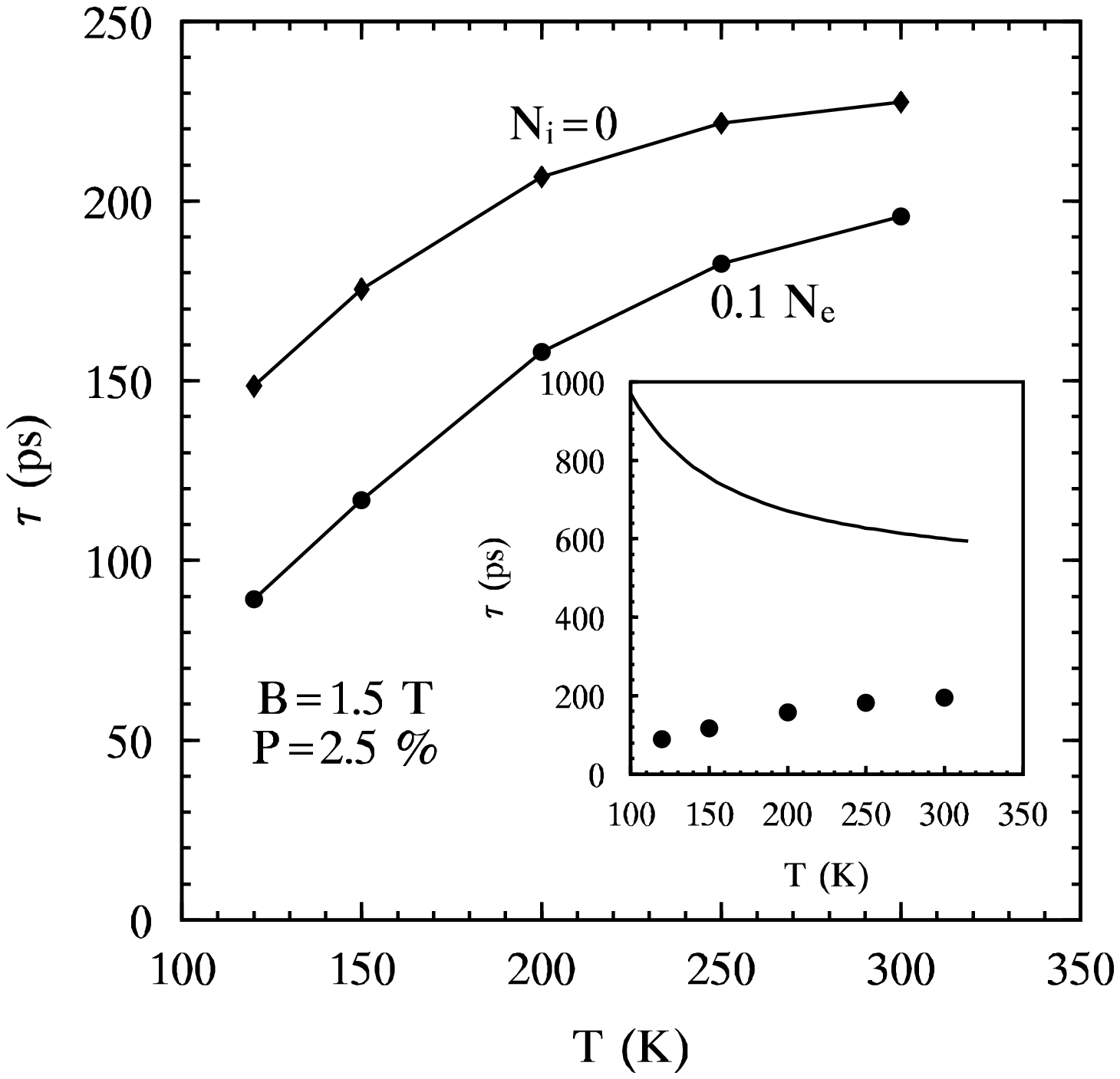,width=9.cm,height=8.5cm,angle=0}
  \psfig{figure=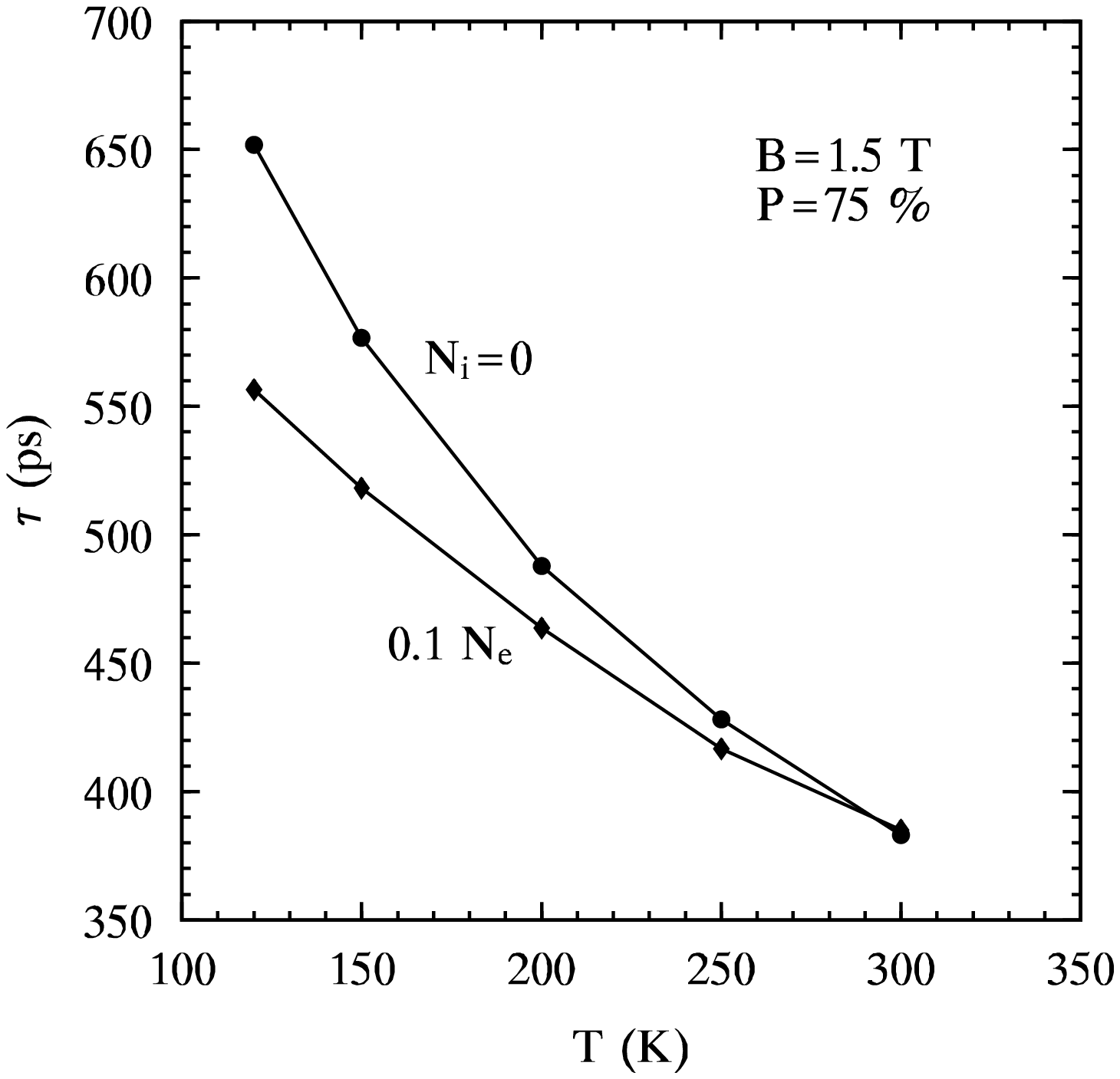,width=9.cm,height=8.5cm,angle=0}
  \caption{Spin dephasing time $\tau$ versus the temperature $T$
    for InAs QW's with 
    spin polarization $P=2.5$~\% (a) and $P=75$~\% (b) 
    under two different impurity levels. 
    Circle ($\bullet$): $N_i=0$; Diamond ($\blacklozenge$): $N_i=0.1 N_e$.
    The lines are plotted for the aid of eyes. 
    The SDT predicated by the simplified treatment of DP term (solid
    curve) and our model (circle) for $N_i=0$ is plotted in the inset
    of (a) for comparison. 
}
  \label{fig5}
\end{figure}

To see more detail of how the spin dephasing depends on the
temperature, we replot
the SDT shown in Fig.~\ref{fig3} as a function of the temperature 
for different impurity levels and different spin polarizations 
in Fig.~\ref{fig5}(a) and (b).
It is seen from the figure that similar to the case of GaAs where the DP term is composed of
the Dresselhaus term,\cite{c0302330} for Rashba term and for low  spin polarization
the SDT also increases  with the temperature for all impurity levels. 
This property is again {\em opposite} to the results of earlier simplified
treatments of the DP effect, where it was predicted that the spin
lifetime decreases with the increase of temperature in the 2D
system.\cite{averkiev,dyakonov}
The SDT based on the simplified model is given
by\cite{wu_jpsj_2001,averkiev,lau} 
\begin{equation}
  \label{eq:taus_ani}
  {1\over \tau} = {\int_0^{\infty} d E_{k} \bigl( 
    f_{k{1\over 2}} - f_{{k}-{1\over 2}}\bigr)\Gamma(k)
    \over \int_0^{\infty} d E_{k} \bigl( f_{k{1\over
    2}}-f_{{k}-{1\over 2}}\bigr)}, 
\end{equation}
in which 
\begin{equation}
  \Gamma(k)= {2\tau_{1}(k)}
  (\alpha_0 E k)^2
  \label{eq:gammak}
\end{equation}
and 
\begin{equation}
  \label{eq:tau_n}
  \tau_n^{-1}(k) =
  \int_0^{2\pi}\sigma(E_k,\theta)[1-\cos(n\theta)]d\theta\ .
\end{equation}
$\sigma(E_k,\theta)$ stands for scattering cross-section.
For comparison, we plot the SDT
predicated by the earlier model and 
by our present many-body theory  in the inset of Fig.\ \ref{fig5}(a). 
>From the inset one can see
that the SDT  predicated by the earlier model is about one order of
magnitude  larger than the one predicated by our theory. 
In the mean time, the SDT
of the earlier mode drops dramatically with the increase of the
temperature. Nevertheless, in our many-body treatment, it rises slightly
with the temperature. 

The giant difference between two models
 lies on the fact  that the earlier simplified model is based on the single particle 
picture  which does not count for the dephasing due to the 
inhomogeneous broadening inherited in the Rashba term,
which is exactly the result of many body effect.\cite{wu_pss_2000,wu_ssc_2002,%
wu_jpsj_2001,wu_epjb_2000,wu_js_2001} 
By comparing the theoretical SDT predicated by the two models, we can
see that the spin dephasing 
due to the inhomogeneous broadening is much more important. In the case we
calculated, the spin dephasing is dominated by the inhomogeneous
broadening. Therefore, it is easy to understand why the earlier
simplified treatment of the DP mechanism gives much slower spin
dephasing. Although there is no experiment results for InAs, 
we have shown that for GaAs the prediction of our many-body theory agrees both
quantitatively and qualitatively with the experiment results.\cite{c0302330}

The temperature dependence of the SDT can be easily
understood when the spin dephasing due to inhomogeneous broadening is
taken into account: When the temperature increases, the inhomogeneous
broadening is reduced as the electrons are distributed to the wider
$k$-states. As a result, the number of electron occupation 
on each ${\bf k}$ state is reduced. It is further
noted that this reduction is mild as function of the
temperature. Therefore, the temperature dependence is quit mild unless
it is within the regime of anomalous peak.

In the region where HF term is important, in addition to the above
mentioned two effects of the temperature acting on the spin dephasing, 
the temperature dependence of HF term should also be taken into
account when we study the temperature dependence of the spin
dephasing. In high spin polarization region, the SDT decreases with the
temperature. However, in the moderately polarized region, 
the temperature dependence of SDT due to the combination of
these three effects is too complicate to be described by a monotonic
function. We replot the SDT as a function of the
temperature in Fig.~\ref{fig5}(b), for the high
 polarization $P=75$~\%, which is near  the anomalous peak
shown in Fig.~\ref{fig3}(a). We can see that due to the reduction of
the HF term, the resonance is removed and the
SDT drops dramatically with the increase of the temperature in the
impurities free sample. 
While for the system with  impurity concentration
$N_i=0.1 N_e$, the HF term is less important than that in the impurities
free sample, and the SDT is less sensitive to the temperature. 
\begin{figure}[htb]
  \psfig{figure=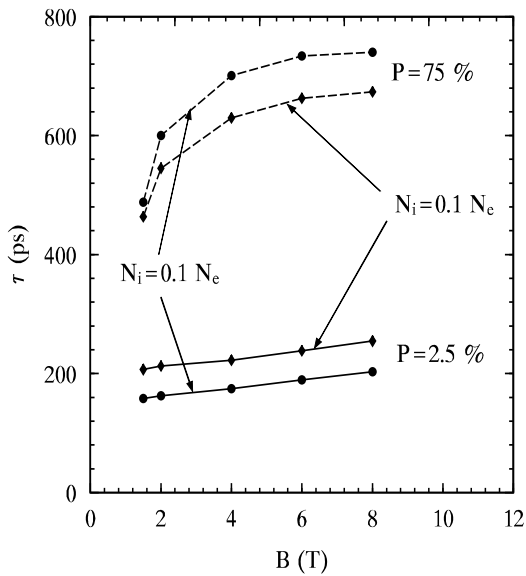,width=9.cm,height=8.5cm,angle=0}
  \caption{Spin dephasing time $\tau$ versus the applied magnetic
    field for InAs QW's for different spin polarizations and different
    impurity levels. 
    Solid curve with dots: $N_i=0$, $P=2.5\ \%$; 
    Solid curve with diamonds: $N_i=0.1N_e$, $P=2.5\ \%$; 
    Dashed curve with dots: $N_i=0$, $P=75\ \%$; 
    Dashed curve with diamonds: $N_i=0.1N_e$, $P=75\ \%$.}
  \label{fig6}
\end{figure}

\subsection{Magnetic field dependence of the spin dephasing}

We now investigate the magnetic field dependence of the spin dephasing. 
In Fig.~\ref{fig6}, we plot the SDT versus the applied magnetic field
for different impurity levels and different spin polarizations. It is
seen that for all the cases we study, the SDT increases with the
magnetic field. This is because
in the presence of a magnetic field, the electron spins
undergo a Larmor precession around the magnetic field. This
precession  suppresses the precession about the effective magnetic field
${\bf h}({\bf k})$.\cite{meier,bronold} 
Therefore the SDT increases with the magnetic field. 
It is pointed out that in 3D electron gas, the magnetic
field also forces electrons to precess around it. This precession
introduces additional symmetry in the momentum space that limits the
${\bf k}$-space available to the DP effect which is anisotropic in
it.\cite{meier,wu_pss_2000,bronold} This can further reduce the spin dephasing. 
However, it is expected that this effect in the 2D case is
less effective than the 3D case as in $z$-direction the momentum
is quantized and the momentum precession around the magnetic field
should be suppressed.  

\begin{figure}[htb]
  \psfig{figure=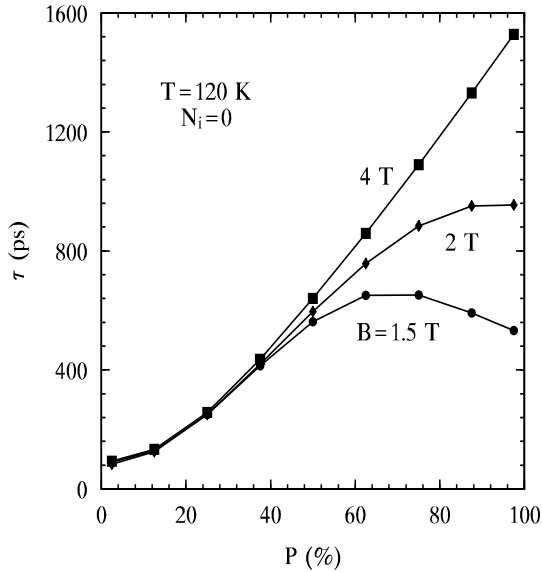,width=9.cm,height=8.5cm,angle=0}
  \caption{Spin dephasing time $\tau$ versus the polarization 
    for InAs QW's at different magnetic field. 
    Circle ($\bullet$): B=1.5\ T; Diamond ($\blacklozenge$): B=2\ T; 
    Square ($\blacksquare$): B=4\ T. The lines are plotted for the aid
    of eyes.
  }
  \label{fig7}
\end{figure}

In additional to the above mentioned effect of the magnetic field on spin
dephasing, one can further see from Fig.~\ref{fig6}, that 
for large polarization, the magnetic
field also enhances the HF term. As we mentioned before,
for large polarization, the contribution from the HF term is
important. 
Increase of the HF term serves as additional
magnetic field which further suppresses the effect of
the DP term ${\bf h}({\bf k})$, and therefore results in a faster rise in
the $\tau$-$B$ curve in the region of $B<4$~T. When the applied
magnetic field exceeds 4~T, the increase of HF term saturates, 
thus the slop of the $\tau$-$B$ curves in the region of $B>4$~T is
reduced to that in the low polarization region. 
To reveal more concrete about the combining effect of the magnetic field and 
the HF term on spin dephasing, we plot the SDT as a function of polarization in
Fig.~\ref{fig7}. It is shown that the rise in the $\tau$-$P$ curve
increases with the magnetic field.
Moreover, the position of the peak in $\tau$-$P$ 
shifts to a larger polarization. This is understood that, it needs a
larger HF term, and hence a larger spin polarization, to achieve the
resonance condition when the magnetic field increases. 
When the magnetic field is raised to 4\ T, it is no longer possible to
form the resonance for all of the polarization. 
As a result the SDT increases uniquely with the polarization and
there is no peak in the  $\tau$-$P$ curve. 

\subsection{Interface electric field dependence of spin dephasing}

\begin{figure}[htb]
  \psfig{figure=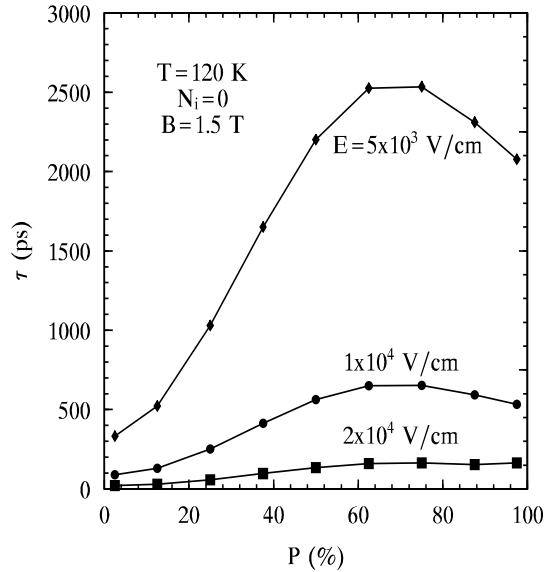,width=9.cm,height=8.5cm,angle=0}
  \caption{Spin dephasing time $\tau$ versus the initial spin
    polarization at $T=120$~K, $B=1.5$~T, $N_i=0$ for three 
    different interface electric fields. 
  }
  \label{fig8}
\end{figure}

We now investigate how the interface electric field affects the spin
dephasing. In Fig.~\ref{fig8} we plot the SDT as a function of the
initial spin polarization for three different interface electric
fields. It can be seen from the figure that when the 
interface electric field decreases from $2\times 10^4$~V/cm to
$5\times 10^3$~V/cm, the SDT increases about 16 folds. It is
understood that when the interface electric field increases, the
Rashba effect is enhanced, and consequently the spin dephasing is also 
enhanced. Moreover, for fixed initial spin polarization, the HF term is reduced
when $E_z$ is increased. As a result,
the interface electric field $E_z$ also changes the anomalous peak in the $\tau$-$P$ curve
as in order to achieve the resonance condition, one has to go to higher
polarization in order to get large enough HF term.
Consequently the resonance peak is smoothed and the position 
moves to higher initial spin
polarization region when $E_z$ increases, which is shown in the
Fig.~\ref{fig8}. 

\section{Conclusion}
In conclusion, we have performed a systematic investigation of the
Rashba 
effect on the spin dephasing of $n$-typed InAs QW's under moderate
magnetic fields in Voigt configuration. Based on the nonequilibrium
Green's function theory, we derived a set of kinetic Bloch equations
for a two-spin-band model. This model includes the electron-phonon,
electron-impurity scattering as well as the electron-electron
interaction. By 
numerically solving the kinetic Bloch equations, we study the time
evolution of electron densities in each spin band and the spin
coherence -- the correlation between spin-up and -down bands. 
The spin dephasing time is calculated from the slope of the envelope of the
time evolution of the incoherently summed spin coherence. We therefore
are able to study in detail how this dephasing time is affected by
spin polarization, temperature, impurity level, magnetic field and
interface electric field. Differing from the earlier studies on spin
dephasing based on the single particle model which only 
considers the effective SF scattering, 
our theory also takes account
of the contribution of many-body effect on 
the spin dephasing.\cite{wu_pss_2000,wu_ssc_2002,wu_jpsj_2001,%
wu_epjb_2000,wu_js_2001} In fact, for the $n$-typed 
semiconductors and the spin polarization studied in the experiments,
this many-body dephasing effect is even more important than
the effective SF scattering as it is one order of magnitude 
larger than the later. 
Equally remarkable is that,
as we include all the scattering, especially
the Coulomb scattering in our many-body theory, now we are
able to calculate the spin dephasing with extra large (up to 100\ \%)
initial spin polarization which is unable to  be calculated by the former single particle theory.

It is discovered that the SDT increases with the initial spin polarization.
Moreover, for low impurity level and low temperature, there is 
a giant anomalous resonant peak in the curve of the SDT versus 
the initial spin polarization. This resonant  peak moves to
high spin polarization and its magnitude is fast reduced (enhanced)
until the whole resonance disappears 
if one  increases the impurity density, the temperature and/or the
interface electric field
(the magnetic field). It is discovered that
this anomalous resonance peak originates from the HF contribution of the
electron-electron Coulomb interaction. 
Under the right spin polarization, the contribution of HF term may reach the
magnitude comparable to the contribution of the DP term as well as the
magnetic field in the coherent part of the Bloch equation and reduces the
anisotropy caused by the Rashba effect---consequently reduces the spin
dephasing. 
As the resonance is the combined effects of the HF
term, the Rashba term and the magnetic field, the magnitude and
position of the 
resonance peak are affected by all the factors 
that can alert the magnitude of the HF term, such as temperature,
impurity scattering, magnetic field as well as the interface electric
field: 
For a given impurity concentration,
when the temperature increases, the HF term reduces. Consequently the
$\tau$-$P$ curve is smoothed and the peak position is moved to higher
spin polarization; For impurity free sample, if the temperature is
raised to 200~K, the HF term is reduced too much to form a
resonance and the anomalous peak disappears; 
The same situation happens when the impurity level increases at
a given temperature as the scattering also lowers the HF term. When the
impurity level is raised to $0.2 N_e$ there is no resonance in the
temperature region we studied; While the increase of the magnetic
field enhances the HF term and results in a faster increase of the
SDT as well as a higher resonant peak in $\tau$-$P$
curve. Moreover, as the magnetic field becomes larger, it needs a
larger HF term and hence a larger polarization in order to achieve the
resonant condition. Therefore the peak position is also moved to
higher polarization; When the interface electric field increases, the
HF term is reduced. Therefore the resonance peak in $\tau$-$P$ becomes
flatter and its position moves to higher spin polarization. 

For low spin polarized regime, the SDT increases when the temperature
rises. This is contrary to the result of earlier simplified
single-particle calculation where the SDT always decreases with the
increase of the temperature. Moreover, the SDT predicted by our many-body
calculation is one order of magnitude faster than the earlier result.
The physics of this feature is due to the additional 
many-body spin dephasing channel due to the 
inhomogeneous broadening provided by the  Rashba term, which by 
combining with the SC scattering also causes spin dephasing.
In the situation we studied, the spin dephasing is dominated by the
many-body dephasing effect.
With the increase of the temperature, the inhomogeneous broadening
reduces and the SDT increases.

In high spin polarization region, 
the contribution of the HF term should be taken into consideration.  
This brings more
complication in the study of the spin dephasing. Usually the SDT can 
not be described by a monotonic function of the temperature and the
impurity concentration in high polarization regime. For
polarization  near the resonance peak in low
temperature and impurities free samples, the SDT decreases dramatically
with the temperature as the resonance is removed when the temperature
increases. Whereas when the impurity concentration is $0.1 N_e$, the
SDT is less sensitive to the temperature. 

As the magnetic field causes the electron spins to precess about it, this
precession will suppress the precession about the effective magnetic
field ${\bf h}({\bf k})$ originated from the Rashba
effect. As a result the spin dephasing is reduced. The magnetic field
also enhances the HF term, which servers as an additional magnetic field
and further suppresses the Rashba effect. Therefore, the $\tau$-$B$ curve
gets a faster increase in the high polarization region.
Our calculation also shows that when the interface electric field
increases, 
the SDT decreases. This is because with the increasing of the
interface electric field, the Rashba term is strengthened. 

In summary we have performed a thorough investigation of the spin dephasing
in $n$-typed InAs QW's. Many new features which have not been
investigated both theoretically and experimentally are predicted in a wide range of
parameters.  

\begin{acknowledgments}
MWW is supported by the  ``100 Person Project'' of Chinese Academy of
Sciences and Natural Science Foundation of China under Grant
No. 10247002. He would also like to thank S. T. Chui at Bartol Research Institute,
University of Delaware for hospitality.
\end{acknowledgments}

\appendix

\section{Appendixes}

\begin{widetext}
The scattering terms of electron distribution functions in the
Markovian limit are given by 
\begin{eqnarray}
  \label{eq:f_scatt}
  {\partial f_{{\bf k},\sigma} \over \partial t}|_{\mbox{scatt}} &=& 
  \biggl\{-2\pi\sum_{{\bf q}q_z\lambda}g_{{\bf Q}\lambda}^2
  \delta(\varepsilon_{{\bf k}}-\varepsilon_{{\bf k}-{\bf
      q}}-\Omega_{{\bf q}q_z\lambda})
  \bigl[N_{{\bf q}q_z\lambda}
  (f_{{\bf k}\sigma}-f_{{\bf k}-{\bf q}\sigma})
  +f_{{\bf k}\sigma}(1-f_{{\bf k}-{\bf q}\sigma})
  -\mbox{Re}(\rho_{{\bf k}}\rho^{\ast}_{{\bf k}-{\bf q}})\bigr]
  \nonumber\\
 && -2\pi N_i\sum_{{\bf q}}U^2_{{\bf q}}
  \delta(\varepsilon_{{\bf k}}-\varepsilon_{{\bf k}-{\bf q}})
  \bigl[f_{{\bf k}\sigma}(1-f_{{\bf k}-{\bf q}\sigma})-
  \mbox{Re}(\rho_{{\bf k}}\rho^{\ast}_{{\bf k}-{\bf q}})\bigr]
  -2\pi\sum_{{\bf q}{\bf k}^{\prime}\sigma^{\prime}}V_{{\bf q}}^2 
  \delta(\varepsilon_{{\bf k}-{\bf q}}-\varepsilon_{{\bf k}}
  +\varepsilon_{{\bf k}^{\prime}}-
  \varepsilon_{{\bf k}^{\prime}-{\bf q}})
  \nonumber\\
  &&\Bigl[
  (1-f_{{\bf k}-{\bf q}\sigma})f_{{\bf k}\sigma}
  (1-f_{{\bf k}^{\prime}\sigma^{\prime}})
  f_{{\bf k}^{\prime}-{\bf q}\sigma^{\prime}}
  +{1\over 2}\rho_{{\bf k}}\rho^{\ast}_{{\bf k}-{\bf q}}
  (f_{{\bf k}^{\prime}\sigma^{\prime}}-
  f_{{\bf k}^{\prime}-{\bf q}\sigma^{\prime}})
  +{1\over 2}\rho_{{\bf k}^{\prime}}
  \rho^{\ast}_{{\bf k}^{\prime}-{\bf q}}
  (f_{{\bf k}-{\bf q}\sigma}-f_{{\bf k}\sigma})\Bigr]
  \biggr\}\nonumber\\
  &&-\{{\bf k}\leftrightarrow{\bf k}-{\bf q},{\bf
  k}^{\prime}\leftrightarrow{\bf k}^{\prime}-{\bf q}\},
\end{eqnarray}
\noindent in which $\{{\bf k}\leftrightarrow{\bf k}-{\bf q},
{\bf k}^{\prime}\leftrightarrow{\bf k}^{\prime}-{\bf q}\}$ stands for the
same terms as in the previous $\{\}$ but with the interchange ${\bf
  k}\leftrightarrow {\bf k}-{\bf q}$ and ${\bf
  k}^{\prime}\leftrightarrow{\bf k}^{\prime}-{\bf q}$. 
The first term inside the braces on the RHS of Eq.~(\ref{eq:f_scatt})
comes from the electron-phonon interaction. $\lambda$ stands for the
different phonon modes, {\em i.e.}, one longitude optical (LO) phonon mode,
one longitudinal acoustic (AC) phonon mode 
due to the deformation potential, and two AC modes due to the
transverse piezoelectric field. $g_{{\bf q}q_z\lambda}$ are the matrix
elements of electron-phonon coupling for mode $\lambda$. 
For LO phonons, $g^2_{{\bf q}q_z \mbox{LO}}=
\{4\pi\alpha\Omega_{\mbox{LO}}^{3/2}/[\sqrt{2\mu}(q^2+q_z^2)]\}
|I(iq_z)|^2$ with 
$\alpha=e^2\sqrt{\mu/(2\Omega_{\mbox{LO}})}
(\kappa^{-1}_{\infty}-\kappa_0^{-1})$. $\kappa_{\infty}$ is the
optical dielectric constant and $\Omega_{\mbox{LO}}$ is the LO phonon
frequency. The form factor 
$|I(iq_z)|^2=\pi^2\sin^2y/[y^2(y^2-\pi^2)^2]$ with $y=q_za/2$. 
$N_{{\bf q}q_z\lambda}=1/[\exp(\Omega_{{\bf q}q_z\lambda}/k_BT)-1]$ is
the Bose distribution of phonon mode $\lambda$ at temperature $T$. 
The second term inside the braces on the RHS of Eq.~(\ref{eq:f_scatt})
results from the electron-impurity scattering under the random phase
approximation with $N_i$ denoting the impurity
concentration. $U_{\bf q}^2=\sum_{q_z}\bigl\{4\pi
Z_i e^2/[\kappa_0 (q^2+q_z^2)]\bigr\}^2 |I(iq_z)|^2$ is the
electron-impurity interaction matrix element with $Z_i$ stands for the
charge number of the impurity. $Z_i$ is assumed to be $1$ throughout
our calculation. 
The third term is the contribution of the Coulomb interaction. 
Similarly, the 
scattering parts of the spin coherence
are given by
\begin{eqnarray}
  \label{eq:rho_scatt}
  {\partial \rho_{{\bf k}}\over \partial t}\left |_{\mbox{scatt}}
    \right . &=&\biggl\{
  \pi\sum_{{\bf q}q_z\lambda}g^2_{{\bf q}q_z\lambda}
  \delta(\varepsilon_{{\bf k}}-\varepsilon_{{\bf k}-{\bf q}}
  -\Omega_{{\bf q}q_z\lambda})
  \bigl[\rho_{{\bf k}-{\bf q}}
  (f_{{\bf k}{1\over 2}}+f_{{\bf k}-{1\over 2}})
  +(f_{{\bf k}-{\bf q}{1\over 2}}+f_{{\bf k}-{\bf q}-{1\over 2}}-2)
  \rho_{{\bf k}}
  -2N_{{\bf q}q_z\lambda}(\rho_{{\bf k}}-\rho_{{\bf k}-{\bf q}})\bigr]
  \nonumber \\
  && + \pi N_i\sum_{{\bf q}}U_{{\bf q}}^2
  \delta(\varepsilon_{{\bf k}}-\varepsilon_{{\bf k}-{\bf q}})
  \bigl[(f_{{\bf k}{1\over 2}}+f_{{\bf k}-{1\over 2}})
  \rho_{{\bf k}-{\bf q}}
  -(2-f_{{\bf k}-{\bf q}{1\over 2}}-f_{{\bf k}-{\bf q}-{1\over 2}})
  \rho_{{\bf k}}\bigr]\nonumber\\
  &&-\sum_{{\bf q}{\bf k}^{\prime}}\pi V_{{\bf q}}^2
\delta(\varepsilon_{{\bf k}-{\bf q}}-\varepsilon_{{\bf
    k}}+\varepsilon_{{\bf k}^{\prime}}-\varepsilon_{{\bf
    k}^{\prime}-{\bf q}})
\biggl(
\bigl(f_{{\bf k}-{\bf q}{1\over 2}}\rho_{{\bf k}}
+\rho_{{\bf k}-{\bf q}}f_{{\bf k}-{1\over 2}}
\bigr)
(f_{{\bf k}^{\prime}{1\over 2}}-
f_{{\bf k}^{\prime}-{\bf q}{1\over 2}}
+f_{{\bf k}^{\prime}-{1\over 2}}-
f_{{\bf k}^{\prime}-{\bf q}-{1\over 2}})\nonumber\\
&&+\rho_{{\bf k}}\bigl[
(1-f_{{\bf k}^{\prime}{1\over 2}})f_{{\bf k}-{\bf q}{1\over 2}}
+(1-f_{{\bf k}^{\prime}-{1\over 2}})f_{{\bf k}-{\bf q}-{1\over 2}}
-2\mbox{Re}(\rho^{\ast}_{{\bf k}^{\prime}}\rho_{{\bf k}-{\bf q}})
\bigr]
- \rho_{{\bf k}-{\bf q}}\bigl[
f_{{\bf k}^{\prime}{1\over 2}}(1-f_{{\bf k}^{\prime}-{\bf q}{1\over 2}})
\nonumber\\ &&
+(1-f_{{\bf k}^{\prime}-{1\over 2}})f_{{\bf k}^{\prime}-{\bf q}-{1\over 2}}
-2\mbox{Re}(\rho^{\ast}_{{\bf k}^{\prime}}\rho_{{\bf k}^{\prime}-{\bf q}})
\bigr]\biggl)\biggl\}
-\bigl\{{\bf k}\leftrightarrow {\bf k}-{\bf q},{\bf
    k}^{\prime}\leftrightarrow
{\bf k}^{\prime}-{\bf q}\bigr\}\ .
\end{eqnarray}
\end{widetext}


\end {document}